# Topology-dependent conjugation effects in graphdiyne molecular fragments

Patrick Serafini[1], Alberto Milani[1*], Matteo Tommasini[2], Carlo E. Bottani[1], Carlo S. Casari[1*]

[1]*Dipartimento di Energia, Politecnico di Milano, via Ponzio 34/3, Milano, Italy*
[2]*Dipartimento di Chimica, Material, Ing. Chimica "G. Natta", Politecnico di Milano, Piazza Leonardo da Vinci 23, Milano, Italy*

*Corresponding authors: Tel: (022399)6392 alberto.milani@polimi.it; Tel: (022399)6331 carlo.casari@polimi.it

**Abstract**

Graphdiynes (GDYs) as two-dimensional carbon structures based on $sp^2$ hybridized aromatic rings connected by sp-hybridized acetylenic linear links are gathering an increasing popularity, both for their peculiar properties and for the promising applications. In these materials, structural features affect the degree of pi-electron conjugation resulting in different electronic and vibrational properties. Indeed, how topology, connectivity between sp and $sp^2$ domains and system size are related with the final properties is fundamental to understand structure-property relationships and to tailor the properties by proper structure design.

By using a computational approach based on density functional theory calculations, we here investigate structure-property relations in a class of 1D and 2D GDY molecular fragments as building block models of extended structures. By analysing how the structure can modulate the pi-electron conjugation in these systems, HOMO-LUMO gap is found to depend on the peculiar topology and connections between linear sp domains and aromatic units. A topological indicator is computed, showing a trend with the gap and with the frequency of the main vibrational mode occurring in Raman spectra. Our findings can contribute to guide the molecular design of new GDY-based sp-$sp^2$ carbon materials, aiming at tuning their properties by precise control of the structure.

Keywords: Density Functional Theory, Raman spectroscopy, electronic properties, aromaticity, graphdiyne, sp-$sp^2$ carbon-based materials

## 1. Introduction

The advent of graphene research has stimulated a strong interest in novel 2D carbon materials based on sp-sp$^2$ hybridization like graphyne (GY) and graphdiyne (GDY). Their ideal properties, predicted from theory, and their synthesis and characterization through a variety of techniques opened to possible applications in different fields of nanoscience and nanotechnology, as discussed in some review papers. [1-15]. In fact, these novel materials show structure-dependent properties, thus representing a library of 2D carbon materials beyond graphene [16-20].

The experimental preparation of these materials is based on the synthesis of extended systems on metal surfaces by organometallic processes involving peculiar molecular precursors [21-23] or is based on rational chemical synthesis of GDY fragments of different types. The latter has been exploited for instance by Haley and co-workers who reported the synthesis of a large variety of GDY fragments [24-28].

GDY crystals share with other polyconjugated materials the large impact structural effects and peculiar topology can have on their properties by strongly modulating the electronic, optical and vibrational response [29-34]. Topological effects in hybrid sp-sp$^2$ carbon materials have been investigated and discussed also very recently [35-36], having a strong relevance for the design, modulation and tuning of the material properties in view of potential technological applications.

In the case of GY, molecular fragments made by aromatic rings connected with monoacetylenic units were theoretically studied for the first time by Tahara and coworkers [37], analyzing the effect of connectivity and topology on electronic and aromatic properties of deydrobenzoannulenes (DBAs). Conversely, in GDYs, the aromatic rings are connected through diacetylenic linkages, resulting in a higher conjugation of the sp-hybridized linear structures. Some open questions regard how topology plays a role in the overall conjugation of these systems and how marked is the size dependence of these effects in GDY fragments of increasing dimension. Indeed, the understanding of the relationship between molecular structure and related electronic/vibrational properties in these materials is important not only for the characterization of fundamental properties of sp-sp$^2$ carbon hybrid materials but also to provide a rationale for the synthesis of new systems with tailored properties.

In this work we aim at investigating the modulation of electronic and vibrational properties of different types of GDY fragments as a function of their structural features and size. Focusing on both 1D and 2D systems, we start first by analyzing molecular chains in which aromatic 6-membered rings are connected by diacetylenic units with para, meta and ortho connections. The effect of topology on the HOMO-LUMO gap was evaluated for increasing length of the 1D systems (i.e. increasing number of molecular units). Then, we analyze the topological dependence of the electronic properties of finite 2D GDY fragments having different numbers of para, meta and ortho connections. On these last fragments, the correlations between electronic properties and topology was analyzed also through the calculation of topological indicators. Focusing on the characterization of sp-sp2 GDY fragments and considering to this aim the relevance of Raman spectroscopy for carbon materials, we also analyze the influence of topological effects on vibrational properties, showing that Raman spectra are affected by these effects in a peculiar way.

## 2. Theoretical Methods

Density Functional Theory (DFT) calculations have been employed to carry geometry optimization, HOMO-LUMO gap evaluation and Raman spectra prediction of a variety of molecular fragments of GDY, where aromatic rings and di-acetylenic units linked together with different type of connectivity (ortho, meta and para conjugation paths, see Figure 1-a). Some of these models were built to consider also the molecules synthetized in previous studies [24-28]. The complete series of all the models here considered are reported in Table S1 of the Supporting Information, together with their HOMO-LUMO gap.

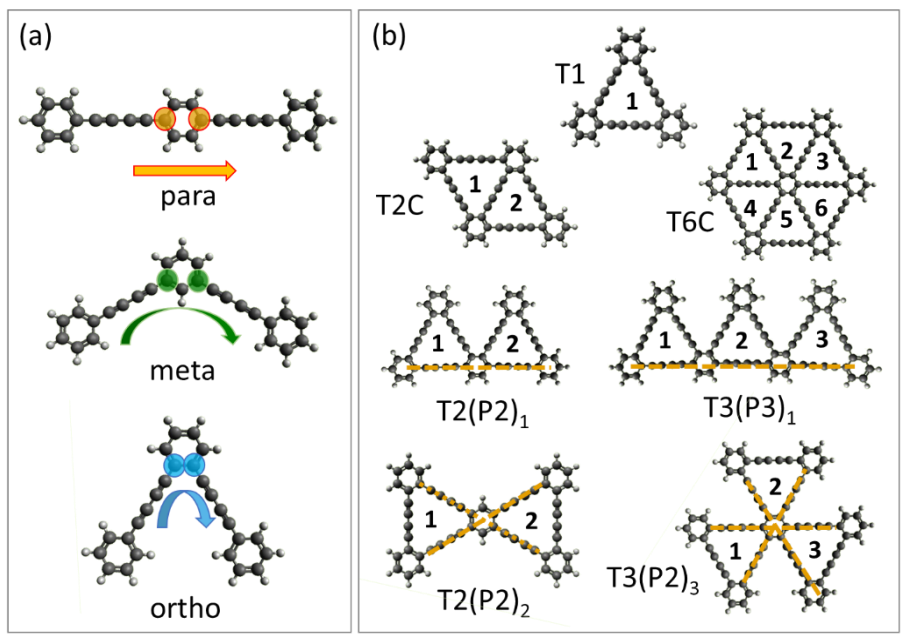

*Figure 1: a) Sketches of para, meta and ortho conjugation paths; b) Sketches of some of the 2D-GDY fragments investigated and related nomenclature. A table reporting the sketches of all the molecular model investigated is reported in the Supporting Information.*

Calculations have been carried at PBE0/6-31G(d,p) level of theory by using GAUSSIAN09 package [38]: this choice has been done also to compare the results found for finite systems with infinite ones. Indeed, one-dimensional polymeric model systems showing different conjugation paths (para-, ortho- and meta) and configurations (cis and trans for meta cases) have been also considered, carrying out periodic boundary conditions DFT calculations at the same level of theory using the CRYSTAL17 code [39-41] to fully optimize their geometry and computed the band-gap. The same computational setup has been adopted in a previous work to study 2D g-GDY and related nanoribbons [36]. For the optimizations of the 1D polymers the tolerance on integral screening (TOLINTEG parameters) have been fixed to 9,9,9,9,80, while the shrink parameters defining Monkhorst-Pack sampling points have been fixed to 50.

All the studied finite and infinite systems were optimized keeping a planar geometry, thus not taking into account the effect of distortion from planarity.

PBE0 functional proved to give a very reliable description in the trends found in the electronic and vibrational properties for a large variety of sp-hybrized carbon systems [42] and this is why it has been adopted also here. It is however well-known that it tends to overestimate the value of the band gap with respect to HSE06 functional [43]: in our previous work [44] the band gap of GDY and its nanoribbons have been computed with both functionals, showing indeed that HSE06 better approaches benchmark values obtained by GW calculations. However, the trends found for both PBE0 and HSE06 with respect to the geometry of GDY nanoribbons are the same, describing the same topological effects on electronic properties.

In order to find structure-property correlations for the analyzed molecules harmonic oscillator model of aromaticity (HOMA) parameter for the aromatic rings is calculated according to the following expression:

$$\text{HOMA} = 1 - \alpha/n * (\sum (R_{opt} - R_i)^2)$$

In this definition, n is the number of bonds taken into the summation and $\alpha$ is an empirical constant ($\alpha$ = 257.7) chosen to give HOMA = 0 for the hypothetical Kekulé structure of an aromatic system and HOMA = 1 for the system with all bonds equal to the optimal value $R_{opt}$ ($R_{opt}$ = 1.388 [45]) $R_i$ are the bond lengths of the CC bonds in the aromatic rings. According to this definition, the limiting cases of graphene and benzene present HOMA values of 0.76 and 0.99, respectively. The value of $R_{opt}$ is calculated according to the following expression:

$$R_{opt} = [R(s) + 2R(d)]/3$$

Where $R(s)$ and $R(d)$ are taken respectively as the carbon-carbon bond lengths of the experimentally measured single and double bonds of butadiene-1,3 molecule in gas phase.
The HOMA values are computed as average values of all the aromatic rings in each structure.
In addition to this parameter, other two indicators were analysed to further investigate these correlations but revealed to be less significant. Related discussion is reported in the Supporting Information (Figure S1).

### 3. Results and Discussion

By analysing a large number of GDY fragments we analyse how topology affects the electronic and vibrational properties of these system, in particular how HOMO-LUMO gap depends on the degree of conjugation, described by the three types of connectivity of the sp-domains to the aromatic 6-membered rings (i.e. para, ortho and meta, see Figure 1.a). Some examples of the investigated 2D GDY fragments are reported in Figure 1-b, together with the nomenclature here adopted to highlight the presence and number of para paths of different lengths. Indeed, the length of the para, ortho and meta pathways is referred to the number of the diacetylenic units that are connected in that connectivity type; the number of carbon bonds in the single units remains unchanged (i.e. four sp-carbon atoms). In the case of the 2D GDY-based molecular fragments, for the number of para, ortho and meta pathways we refer to the multiplicity of the different pathway-types in the structures. Starting from the fundamental unit characterized by a "triangular ring" named T1 the nomenclature TXC or TX(PY)$_z$ indicates how many of these rings are present (X), if they are condensed (C, that is if they share one diacetylenic linkage), which is the length (Y) of the para-conjugated paths that they form (i.e. how many diacetylenic units are connected in para-position with respect to the aromatic rings) and how many of these paths are present (Z).

### 3.1 Gap modulation in 1-dimensional molecular fragments

We start by considering one-dimensional models as the simplest systems to investigate the extent of conjugation across the different path types as a function of the dimension of the fragment. Starting from the monomeric building block (i.e. a di-phenyl polyyne with 4 sp-carbon atoms [46] oligomers of increasing length and connected in para, ortho and meta (cis and trans) position were studied and compared to the infinite polymer described as a 1D crystal in periodic boundary conditions.
The HOMO-LUMO gap as a function of the fragment length is reported in Figure 2 for all the possible configurations (the numerical values of the gap are reported in Table S1 of the Supporting Information). The gap values of 2D fragments composed of "triangular" units (each one formed by three sp-chains and aromatic rings) are reported for comparison, up to the infinite limit which corresponds to the smallest GDY nanoribbon. This aims at introducing also the effect to the two-dimensional extension of the fragments with respect to simple one-dimensional conjugation represented by the other models.

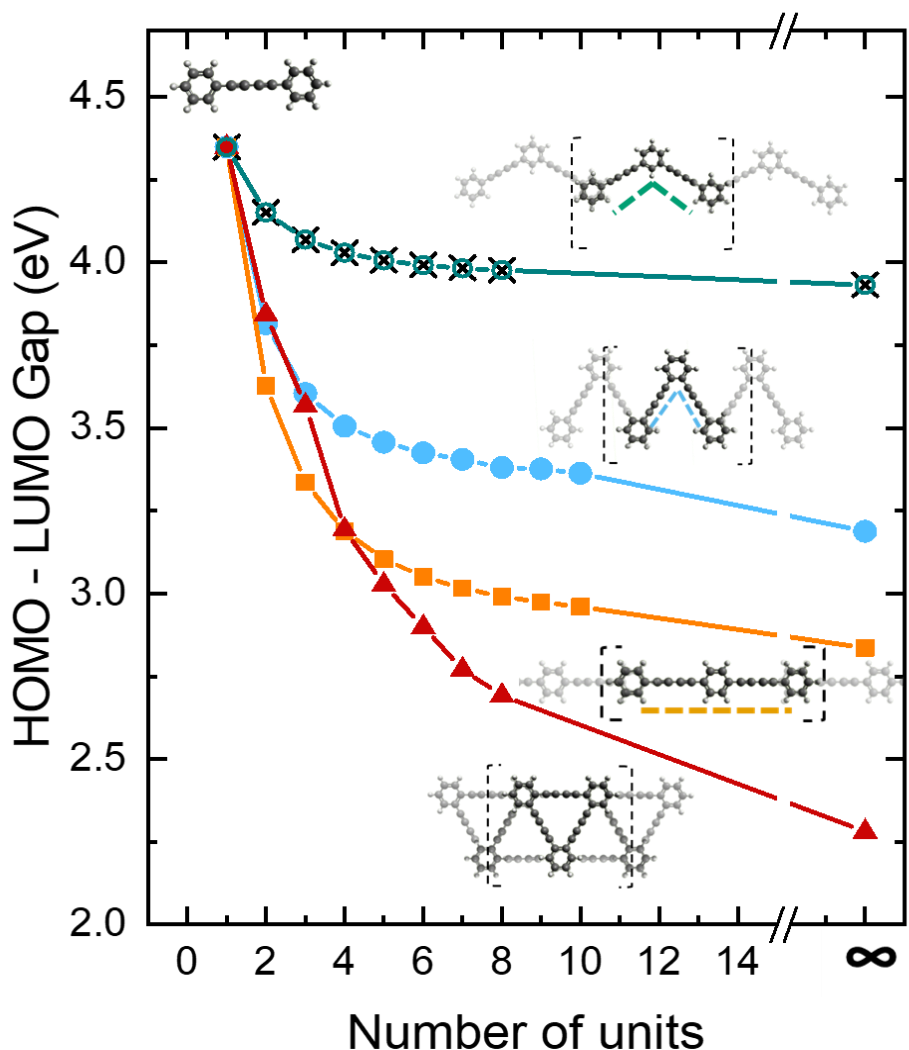

*Figure 2:* *Modulation of the DFT-computed (PBE0/6-31G(d,p)) HOMO-LUMO gaps (in eV) of one dimensional fragments (up to the 1D infinite polymer) formed by connecting N diacetylenic units in ortho-, para-, metha- positions or forming triangular rings. The single T1 ring corresponds to a value N=2 on the x-axis and T2C to N=3, the related sketches are reported in the figure. In the case of meta-connected units, both cis- and trans- configurations have been considered but the associated values are superimposed thus not showing any configuration-dependent effect on the gap. The ∞ symbol stands for the infinite polymeric 1D case, whose gap value has been calculated with periodic boundary conditions Density Functional Theory simulations at the same level of theory (PBE0/6-31G(d,p))*

In all the cases a general decrease of the gap is observed when increasing the number of units. When moving from the monomer to the oligomer with two diacetylenic units, it is clear that the conjugation is larger (gap is lower) for para-conjugated units (gap drops from 4.35 to 3.63 eV moving from N=1 to N=2), followed by ortho-conjugated units (3.81 eV) and finally by meta-conjugated fragments with a gap exceeding 4 eV (4.15 eV). By increasing the number of units in the case of meta-conjugation the trend rapidly converges with no difference for trans or cis configuration. Meta-conjugation strongly limits the extent of conjugation across aromatic 6-membered units, thus preventing the modulation of the gap to lower values. Conversely, para and ortho connections can promote a significant pi-electron conjugation across different units and a saturation effect is observed in both cases only for 1D fragments longer than six or seven units. Para connections are the most effective ones in promoting a larger conjugation, with a HOMO-LUMO gap rapidly dropping to values below 3 eV. This is in agreement to what is observed in similar structures in which diacetylenic linkages are substituted by acetylenic ones [37]. The infinite polymers reveal indeed a band gap of 3.188 eV for the ortho case and of 2.836 eV for the para- one, both much lower than the value of 3.933 eV predicted for the meta-conjugation (for both cis and trans configurations).

Similar to the modulation of the gap, BLA values on the diacetylenic chains could be expected to be modulated by the different extent of pi-electron conjugation. However, the modulation of the BLA is negligible. In fact, moving from the monomer to the two-unit oligomers the BLA value changes from 0.1387 Å to 0.1374 Å for para-connection, to 0.1376 Å for ortho-connections and 0.139 Å for meta-connections revealing that conjugation is almost prevented for meta-, while it is somehow relevant for para- and ortho- cases (Figure S2). In the two latter cases as soon as the oligomer length increases, no further relevant changes of BLA are predicted (in the para-conjugated polymer it is 0.1363 Å and in the ortho one 0.1368 Å), thus revealing that the modulation of the gap is not directly related to the modulation of the bond lengths in the sp-carbon domains. Indeed, carbon-carbon bonds in the diacetylenic linkages remain practically unchanged with the increasing of the oligomer length.

The case of oligomers formed by connected triangular units reveals further effects. For N=2, the triangular structure is formed by only one triangular unit (T1) (see Figure 2) in which the longest path consists in two diacetylenic units connected in ortho- position. In this fragment the gap (3.84 eV) equals the case of the ortho-oligomer having two units (3.81 eV). Moving to two condensed triangular units (T2C in Figure 1-b, N=3 in Figure 2), the longest path is now formed by three ortho-conjugated diacetylenic units and a close correspondence in the gap (3.57 eV) is found with the ortho-oligomers of three units (3.60 eV). However, increasing again the number of condensed triangular units, a very peculiar effect is found. For N=4 three triangular units are connected (T3C) (see Table S1) forming both one ortho-path of four diacetylenic units and one para-path of two units. In this case, the gap (3.19 eV) is lower than the corresponding paths taken as single fragments (i.e. 3.51 eV and 3.63 eV, respectively). Increasing further the number of condensed triangular units, the HOMO-LUMO gap rapidly decreases at a much lower values than those predicted for the infinite para-conjugated polymer. This indicates that the length and type of the longest one-dimensional path in the fragment is not enough to describe the trend in the gap. Conjugation affects the gap in a non-trivial way with more complicated conjugation paths play an effective role. In fact, in condensed triangular units of increasing size both para- and ortho-conjugated paths of different lengths can be identified, but the predicted gap is very different from the one computed for the individual case of an ortho/para oligomer of the same length. This suggests the presence of combined effects involving both ortho and para paths that cannot be separated when larger fragments are considered. The band gap observed for the 1D system of connected triangular units (i.e. the smallest GDY nanoribbon, see also Ref. [44]) presents a computed band gap value of 2.281 eV, well below the para-polymer, demonstrating again the importance of conjugation paths interacting in a 2D space.

The present discussion based on a bottom-up approach becomes quite straightforward when adopting a top-down perspective, starting from the 2D infinite GDY crystal and introducing a confinement in one direction to form GDY nanoribbons (GDYNR) of decreasing width. We have reported a full computational investigation of the band gap modulation of GDYNR [44]. In that work we showed that the band gap of 1.629 eV for the 2D GDY crystal progressively increases for GDYNR of decreasing width reaching the highest value of 2.281 eV in the smallest width nanoribbon reported in Figure 2 for N=∞ (yellow line). This modulation has been already demonstrated since the first investigations of GDYNR [47-49] and parallels what is also observed in graphene nanoribbons. However, in finite-length oligomers, different two-dimensional conjugation effects are present. This will be elucidated in the following section focused on the topological effects in 2D fragments.

### 3.2 Gap modulation in 2-dimensional molecular fragments

Starting from the results reported in the previous section, molecular fragments extending in two dimensions are discussed here, considering in particular the different effects driven by para-, ortho- and meta paths. In sp-hybridized linear carbon system, the bond-length alternation (BLA) usually shows a very nice correlation with both the band gap and the vibrational frequency of longitudinal normal modes [50-54]. Conversely, in hybrid sp-$sp^2$ carbon GDY materials the BLA of di-acetylenic linkages is not significantly affected, in agreement with the case of GY fragments [37]. Substantial geometrical changes on bond lengths are found on the aromatic units connected by these di-acetylenic fragments in the different paths. In fact, carbon-carbon bonds of the $sp^2$ hybridized carbon atoms in the aromatic units are strongly affected by the different connections of the sp domains to the $sp^2$ ones. On this basis, as a structural index to correlate the topology with the computed HOMO-LUMO gap we evaluated the harmonic oscillator model of aromaticity (HOMA) parameter for the aromatic rings, calculated as suggested in Ref. [37] and as defined in the original work by Krygowski et al. [45] (see Theoretical method section).

HOMA is shown to decrease by increasing the number of triangular rings (18-membered rings) in the molecular fragments due to the elongation of conjugated bonds. Hence this parameter is able to describe all the topological effects which affects the structure of the aromatic ring, including conjugation effects responsible for modulating the HOMO-LUMO gap of the systems. Therefore, HOMA is a reliable index to show a correlation with the electronic gap, allowing to investigate relevant trends in topology-vs-gap relationship.

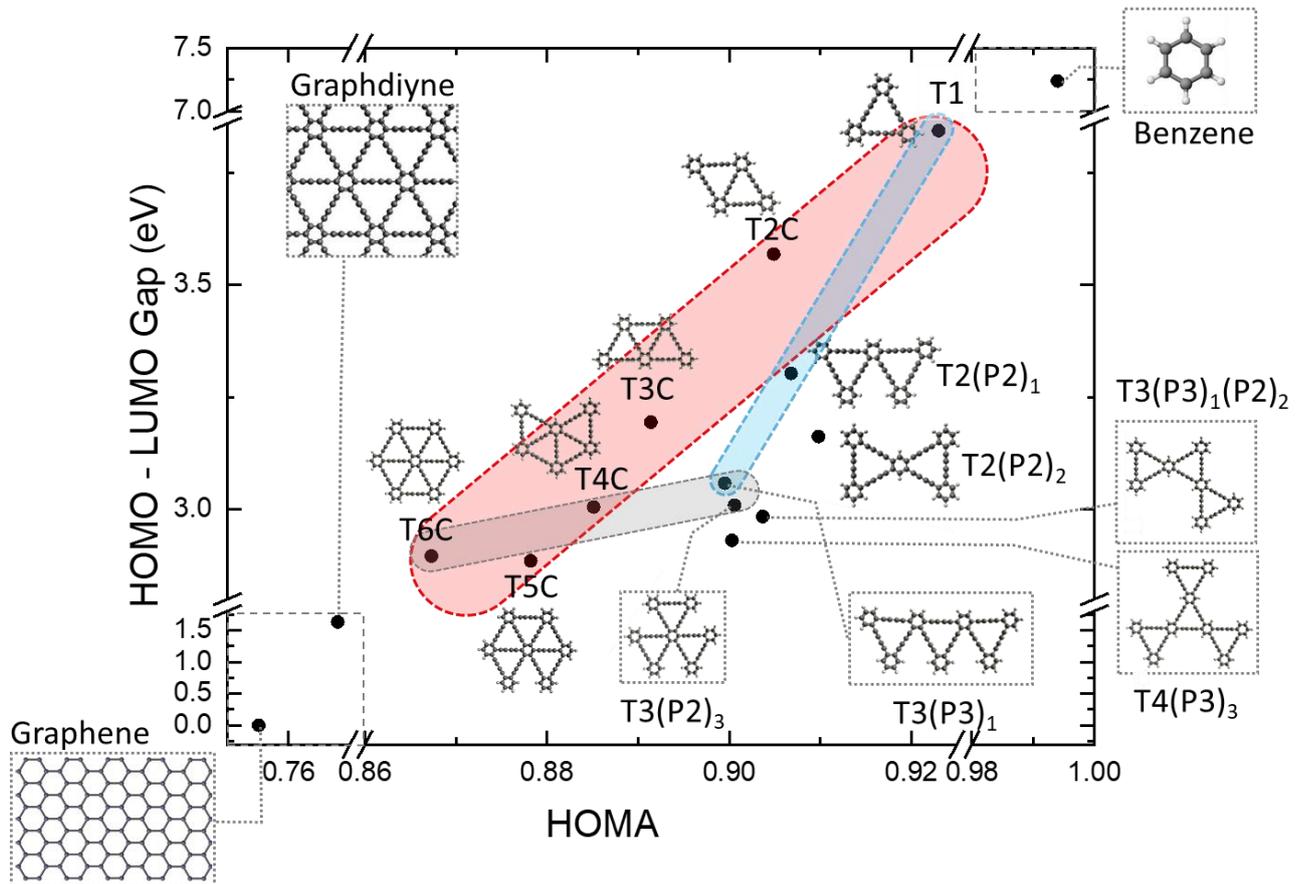

**Figure 3:** *Correlation between computed HOMO-LUMO gaps (in eV) and mean HOMA values of some of the 2D-GDY fragments investigated (see Table S1). Shaded areas show some peculiar trends discussed in the text. Graphene and benzene are reported as a reference. See the methods section for a definition of the HOMA parameter.*

As a reference we report graphene (i.e. completely fused aromatic rings) showing zero band gap and the lowest HOMA (0.755) amongst pure carbon systems. The upper limit is benzene with a gap of 7.24 eV and HOMA=0.99, considered as the isolated aromatic units and thus presenting the higher gap correlated with the higher HOMA. All the systems here investigated can be viewed as aromatic rings interconnected in different geometries thus forming GDY fragments of different sizes whose HOMA values falling between these two limits (see Figure 3).

A general trend indicates smaller gap for smaller HOMA. Interestingly, considering only sp-sp$^2$ structures the two limits are represented by the single triangular molecule (T1) with a HOMO-LUMO gap of 3.84 eV and HOMA value of 0.92 and the infinite 2D g-GDY crystal with a band gap of 1.63 eV HOMA value of 0.77. g-GDY has larger gap and HOMA values than graphene but lower than all the finite GDY fragments showing that HOMA is a reliable index to correlate the electronic gap of these sp-sp2 carbon systems with the degree of pi-electron conjugation modulated by structural changes induced by topology.

A clear relationship can be found by considering the subgroup of structures of condensed triangles of increasing number from the single triangle to the 6 condensed one (i.e. T1, T2C, T3C, T4C, T5C and T6C, highlighted in red). A progressive decrease of the gap is correlated to the decrease of the HOMA with increasing the number of fused triangular rings (that share at least one side of the triangular unit), which consist respectively of 1,2,3 4 and 6 triangles in these five structures. The first three cases correspond to the oligomers of the GDYNR already discussed in the previous section. In fact, moving from the triangular ring T1 to T2C an ortho-path of three diacetylenic units is formed, which is responsible for the decrease of the gap and of the HOMA. In T3C the GAP and HOMA are further lowered due to the dominating effect of para-conjugation, having a para-path of length 2 (i.e. made by 2 di-acetylenic units). T4C has a lower gap due to the increase in the number of these paths, which are now 2 of length 2, while T5C and T6C have the lowest gap values, presenting both 3 para-paths of length 2.

A similar trend with a different slope is found by considering an increasing number of triangular units linearly connected by sharing only one corner an namely starting from T1 and then considering T2(P2)$_1$ and T3(P3)$_1$ ( Figure 3, highlighted in light blue). The observed decrease of gap and HOMA with increasing the number of units can be related to the progressive lengthening of para-pathways respectively of length 2 and 3 in T2(P2)$_1$ and T3(P3)$_1$. However, the comparison between T2(P2)$_1$ and T3C reveals what happens in very similar structures both having one-para path of length 2. The lower gap in T3C results from the "closure" of the middle ring to form a system of three-fused ring instead of only two. The fact that the trends highlighted in red and in light blue respectively in Figure 3 are slightly different can be related to the HOMA definition and its relationship to the structure topology. Indeed, in the fused structures highlighted in red the increase of the number of aromatic rings having more than two interconnected di-acetylenic bridges in ortho-position, affects the CC bond lengths in the aromatic rings. On the other hand, in the open structures highlighted in light blue in Figure 3, most of the aromatic rings are connected only through two di-acetylenic chains connected in ortho-position, and this topology will even more significantly affect CC bond lengths in the rings, differently from the closed fragments. This subtle topological dependence of individual CC bond-lenghts is therefore taken into account in the HOMA value.

Another interesting effect related to the type of connection between only two triangular rings can be understood looking at T2C, T2(P2)$_1$ and T2(P2)$_2$ . T2C has no para paths and a larger gap, while the the latter two structures have two triangular units connected at one corner but forming two different path of length 2 in T2(P2)$_2$ instead of only one in T2(P2)$_1$ .As expected, due to the increase in number of para-paths, T2(P2)$_2$ has a lower gap. Such structures show a decreasing gap at a substantailly fixed HOMA (it shows only a slight increase). This behaviour can be disussed by the inspection of the topology of these structures. While in T2(P2)$_1$ all the aromatic units have multiple diacetylenic chains in ortho-position, in T2(P2)$_2$ the central aromatic ring possesses four diacetylenic bridges where two ortho- and two meta- connection can be identified.
Differently than peripheral ortho-connected rings, the presence of meta-connections (which interreupt conjugation) do not affect the CC bond-leghts of this rings and the interplay between these ortho and meta-connections turns out to affect only slightly the HOMA value.
A further trend related to the existence of para-pathways is revealed by comparing T3(P3)$_1$ T4C, T3(P2)$_3$ and T6C (Figure 3, highlighted in grey). The latter three models possess only para-paths of length 2, which are 2 in number for T4C and 3 for T3(P2)$_3$ and T6C, while T3(P3)$_1$ has only one para-path of length 3. Despite the longest path, the gap of T3(P3)$_1$ is larger than the other three, suggesting that a larger number of shortest path can decrease the gap more than the length of the para-conjugated segments. The case T4C further shows that even if only two para-paths of length 2 are present, by maximizing the number of fused triangular rings, conjugation will dominate over a larger length of the para-path. Considering now T3(P3)$_1$(P2)$_2$ and T4(P3)$_3$ (see Figure 3 and Table S1), where now we have respectively one para-path of length 3 plus 2 para-path of length 2 in T3(P3)$_1$(P2)$_2$ and 3 para-path of length 3 in T4(P3)$_3$, a lower gap is found with respect to the previous models but they are both slightly larger than T6C (3 para-path of length 2 where the number of fused triangular rings is maximized). Very relevant is the comparison between T6C and T4(P3)$_3$: T6C has three p-path of length 2 while T4(P3)$_3$ has three para-path of length 3. The latter should have a lower gap based on the assumption that the length and the number of para-paths mostly affect the gap value. However, they both present a very similar gap value, demonstrating that the formation of a multiply-fused structure has a dominating effect in reducing the band gap with respect to a more "open" one (with longer para-path).
Our analysis reveal a subtle interplay of different topological effect in determining the degree of conjugation and hence the HOMO-LUMO gap in GDY fragments. We have shown that para pathways dominate over ortho and meta ones. The number of para pathways dominate over the length of each path and for the same number of para paths fused triangular structures showed a larger conjugation (i.e. smaller HOMA).

### 3.3 Modulation of the relevant marker bands in Raman spectra

The very intense Raman band related to longitudinal normal modes of vibration (collective CC stretchings on the linear sp-domains, Effective Conjugation Coordinate - ECC mode [50-54]) reveals a significant modulation for the different structures and it is another important experimental observable which could help in establishing structure-properties relationships in GDY-based system.

For this reason, we calculated the Raman spectra of the investigated fragments to correlate electronic and vibrational properties and to assess the use of spectroscopic measurements to identify the presence of peculiar topologies (see Figure 4). No clear correlation with the peculiar structure of the different fragments is found in the aromatic ring vibrations below 1700 cm$^{-1}$. Hence we focus on the main Raman active bands associated to stretching vibration of sp-carbon atoms falling in the 2000 and 2400 cm$^{-1}$ spectral region which is a peculiar and intense marker band of sp-carbon systems. The normal mode related to this band is known as Effective Conjugation Coordinate (ECC) mode and consists in simultaneous lengthening of the triple bonds and shrinking of the single ones along the sp-carbon based unit. This band is strongly related to pi-delocalization along the sp-carbon domains and it is usually modulated by the BLA.

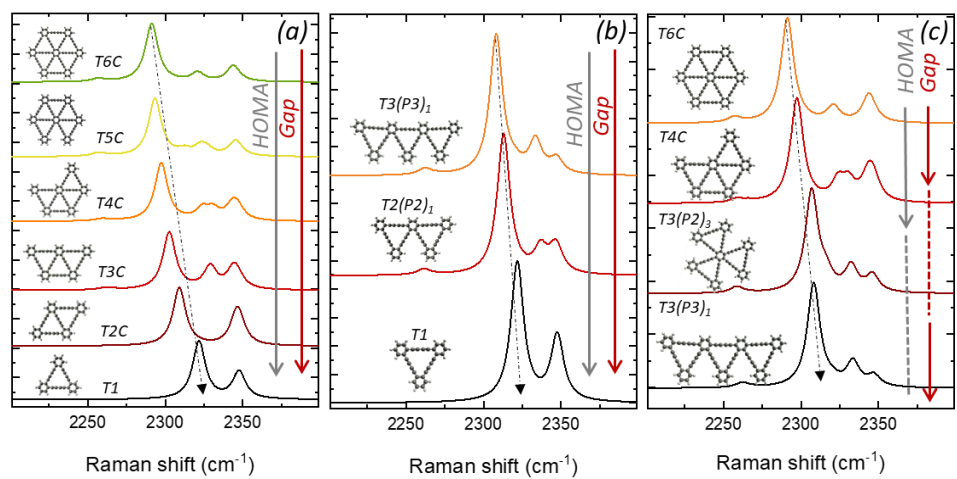

*Figure 4:* *DFT-computed Raman spectra of 2D-GDY molecular fragments for different subgroups of structures. The corresponding trend of HOMA and gap is represented by arrows (dotted line indicates no substantial variation).*

As shown in Figure 4.a, in fragments with increasing number of fused triangular rings the most intense ECC mode shows a progressive red shift in the frequency (more than 30 cm$^{-1}$). This redshift is accompanied by a decrease of both gap and HOMA.

A similar trend is observed when increasing the length of a single para path as in T1, T2(P2)$_1$ and T3(P3)$_1$. The ECC mode frequency changes from 2322 cm$^{-1}$ to 2308 cm$^{-1}$ for increasing length of the para-conjugation pathway and with a parallel decrease of gap and HOMA (Figure 4.b). More complex is what happens in Figure 4c for T3(P3)$_1$, T3(P2)$_3$, T4C and T6C. T3(P2)$_3$ and T4C. They have the same gap (3.009 and 3.005 eV respectively) but the main Raman peak of T4C is red-shifted in frequency with respect to T3(P2)$_3$ (10 cm$^{-1}$, from 2307 to 2297 cm$^{-1}$). This is correlated to the larger number of fused 6-membered rings that are observed in T4C. Moreover, T3(P3)$_1$ (band gap of 3.06 eV) has an intense Raman active peak whose normal mode is located at the same frequency of the main Raman active mode of T3(P2)$_3$, consistently with the very similar gap and the same HOMA values. Finally, the most intense peak associated to the ECC mode of T6C reveals to be the most red-shifted between the others (2291 cm$^{-1}$), again in very good agreement with the trends observed in the gap values.

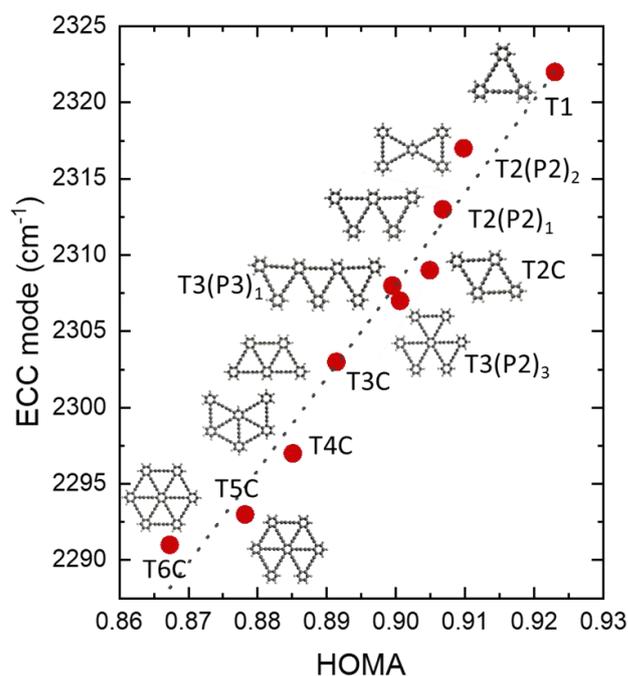

*Figure 5: Correlation between HOMA and ECC mode wavenumbers*

To summarize these trends, we report the HOMO-LUMO gap as a function of the vibrational frequency in Figure S3 of the supporting information.

An impressive correlation is found between HOMA and ECC wavenumbers, with a relationship which is very close to linearity, as shown in Figure 5. This indicates a close relation between the structural variations induced on the aromatic rings by the possible different topology and the vibrations of the linear sp-domains. In particular, we underline that sp-carbon chains vibrations are indirectly affected by geometry variations on the hexagonal rings, indicating further how these rings are the key connecting units in modulating the extent of pi-electrons delocalization in these hybrid sp-sp2 carbon materials. Therefore, the HOMA parameter turns out to be significant to investigate not only the trends in the electronic gap, but also in the vibrational features of these systems. In sp-carbon structures, the correlation between structural/vibrational/electronic properties is usually based on the BLA. In fact, BLA is a structural index which directly modulates the frequency of the ECC mode and is directly influenced by the degree of conjugation. In hybrid sp-sp2 carbon materials studied here, the BLA predicted on the diacetylenic domains does not show remarkable difference. This means that conjugation has a more complex effect on the structural parameters responsible for modulating the gap and the Raman frequency of the ECC mode. It is indeed the HOMA index that is able to describe both the different gap and the variation in the ECC mode frequency as a function of the different topology of the GDY fragments.

### 3.4 Synthesis approaches

Concerning the synthetic approaches which could be employed to producethe GDY-based molecular fragments here discussed, different strategies have been already implemented and reported in literature. The first one is based on a bottom-up approach by chemical synthesis that have been explored with different techniques in several studies [24-28,55-57]. Indeed, the Haley group in particular was able to synthesize various molecular fragments using inter- and intra-molecular cyclization approach based on Sonogashira cross-coupling reactions between precursors with alkynylated groups, thus obtaining acceptable overall yields and easy product isolation and purification. In this context, the synthesised molecules reported as **14**, **15**, **19**, **20**, **21**, **22**, **23**, **26**, **29** in [26] corresponds indeed to the structures T0(P2)$_3$ (Table S1), T1, T2(P2)$_2$, T2(P2)$_1$, T3C, T2C, T3(P2)$_3$, T3(P3)$_1$(P2)$_2$, T4(P3)$_3$ reported in this paper. For some of the synthetized molecules that have been also investigated in our work, we here compare the predicted HOMO-LUMO gap with the measured lowest energy peak of absorption [55-56]. As can be seen in Table 1, there is a good agreement between the computed HOMO-LUMO gap and the lowest energy absorption peaks reported in literature [55-56]; this comparison further confirm the peculiar physicochemical phenomena described in this work.

*Table 1:* *Theoretically computed HOMO-LUMO gap and experimental lowest energy absorption peaks are compared for several synthesized structures*

| Structure | HOMO-LUMO gap (eV) | Lowest energy absorption peak (eV) |
|---|---|---|
| T1 | 3,84 | 3,36 |
| T2C | 3,57 | 3,02 |
| T2(P2)$_1$ | 3,3 | 2,95 |
| T3C | 3,19 | 2,91 |
| T2(P2)$_2$ | 3,16 | 2,88 |
| T3(P3)$_1$(P2)$_2$ | 2,98 | 2,73 |
| T4(P3)$_3$ | 2,93 | 2,77 |

A second possible technique that have been used for the preparation of these molecular fragments is the on-surface synthesis. Indeed, the rapidly growing of this technique has demonstrated tremendous potential in the high-precision bottom-up construction of low-dimensional carbon nanostructures. Its great advantage relies on fostering selective chemical reactions between molecular precursors on metal surfaces under ultra-high vacuum (UHV) conditions. To this aim, molecular precursors, functionalized with alkynyl halides, are designed to favour the adsorption and the subsequent on-surface dehalogenative/dehydrogenative homocoupling reaction, where the substrate acts as a catalytical template triggering the reaction in mild conditions [22, 33-34]. In this way, acting on the precursor, it is possible to modulate precisely the synthesis in order to obtain GDY-based materials with peculiar topologies and dimensions.

Starting from the already synthesized systems and applying further these approaches, we think that it could be possible to synthetised all the GDY-molecular fragments investigated in this paper, focusing also on Raman spectroscopy, here proved to be very sensitive to the peculiar topology of these systems.

## 4. Conclusions

Hybrid sp-sp$^2$ carbon system belonging to the graphdiynes family are nowadays among the main actors at the frontier of the research on nanostructured carbon materials. Most of the efforts in this context have been driven up to now to the synthesis on new structures, producing a large amount of both finite-dimension or extended structures. The possibility to play with topology, connectivity or sp-sp$^2$ carbon ratio allows to create an endless number of different structures, potentially characterized by different electronic/vibrational properties and therefore potentially allowing to tune the final response of the material by a proper molecular design. In our work we aimed at giving a contribution in the understand of structure/properties relationships of GDY materials, investigating the connection between the molecular structure of some fragments and their fundamental physicochemical properties. Indeed, HOMO-LUMO gap, molecular structure/topology and Raman response of GDY fragments have been shown to display some peculiar and general trends, giving useful information for the design of new promising systems. One-dimensional models allow to verify the dependence of the gap on the type of conjugation path (para, ortho or meta) where para-path shows the largest conjugation. By looking at the different trends highlighted in Fig. 3, it is possible to assess that the length and number of para-pathways significantly modulate the HOMO-LUMO gap among the other possible conjugation pathways (ortho and meta), in agreement with what observed in GY [37]. However, we found that the number of "condensed" triangular units can be even more important in reducing the gap. Despite these effects, the gap appears to reach a stable value as the number of "triangular" units is increased in the fragments, suggesting that to approach the limit of infinite 2D GDY (band gap of 1.629 eV [44]) very large fragments or large GDYNR are required, thus possibly making the use of bottom-up chemical synthetic routes unfeasible. The Raman frequencies associated to vibrational stretching modes localized on the sp-carbon domains show a correlation with the predicted HOMO-LUMO gaps. In particular, a red shift of the frequency is found for systems possessing a lower gap. These correlations can be used, at least qualitatively, to evaluate the modulation of the gap for molecules characterized by different structures and topologies possibly in relation to spectral measurement of previously synthetized systems.

Our results unveil the subtle interplay between structure, topology and the electronic/vibrational properties of 2D hybrid systems based on sp and $sp^2$-hybridized carbon and offer some insight for engineering new materials with tailored properties.

## Acknowledgements

Authors acknowledge funding from the European Research Council (ERC) under the European Union's Horizon 2020 research and innovation program ERC—Consolidator Grant (ERC CoG 2016 EspLORE grant agreement No. 724610, website: www.esplore.polimi.it)

## Vitae

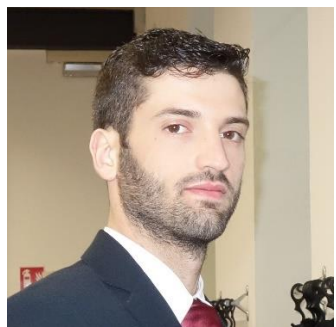

**P. Serafini**

Patrick Serafini was born in 1994 in Teramo, Italy. He is a PhD student in Energy and Nuclear Science and Technology since 2019. He graduated magna cum laude in Materials Engineering and Nanotechnology in December 2018 with a thesis on Density Functional Theory simulations of two-dimensional sp-$sp^2$ carbon nanostructures. His interests are mainly focused on molecular modeling of sp-carbon based nanostructures and nanomaterials. PhD project: theoretical investigation of the electronic, vibrational and optical properties of carbon atom wires (CAWs) and hybrid sp-sp2 nanostructures by means of Density Functional Theory (DFT).

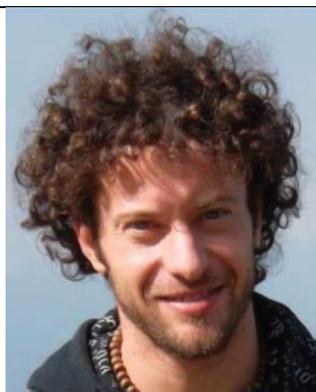

**A. Milani**

Alberto Milani was born in 1980 in Erba – Italy. In 2005, he graduated in Materials Engineering at Politecnico di Milano where he also obtained a PhD in Materials Engineering in April 2009. Since 2005, he joined the research group led by Prof. Zerbi and Prof. Castiglioni as assistant professor and researcher and from 2016 he joined the research group of Prof. C.S. Casari. His scientific interest focuses on computational materials science and vibrational spectroscopy, in particular on the application of state-of-the-art quantum chemical methods for the investigation of the spectroscopic response of polyconjugated molecules and polymers

| 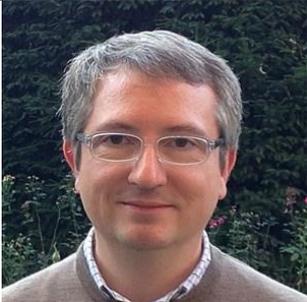 **M. Tommasini** | Matteo Tommasini was born in 1972 in Desenzano del Garda (Brescia, Italy). At Politecnico di Milano (Italy) he graduated in Nuclear Engineering in 1998 and he obtained his PhD in Materials Engineering in 2002 under the guidance of Prof. G. Zerbi and Prof. C. Castiglioni. He is associate professor in Materials Science and Technology at Politecnico di Milano. His research interest focuses on the applications of Vibrational Spectroscopy and Quantum Chemistry to Materials Science and Biotechnology. |
|---|---|
| 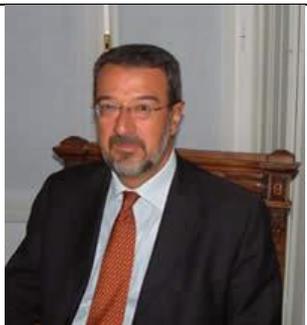 **C. E. Bottani** | C. E. Bottani was born in 1951. He graduated in Nuclear Engineering at Politecnico di Milano summa cum laude in november 1975. Since March 1st 2000 he is full professor of Condensed Matter Physics at Politecnico di Milano. His Research activity is mainly focused on: phonon spectroscopy, cluster and film deposition by pulsed laser ablation and Nanophotovoltaics. He is author of 245 publications among which several monographic chapters in international volumes and five patents. Two papers are in co-authorship with the Nobel laureate Richard Smalley. He has been the Scientific Coordinator of the Center of Excellence "Engineering of Nanostructured Materials and Surfaces" of Politecnico di Milano. |
| 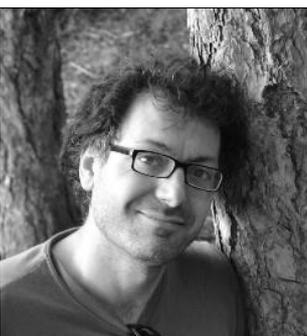 **C. S. Casari** | C. S. Casari was born in Milan in 1974, he got a Master degree in Electronic Engineering in 1999, had research collaborations at Politecnico di Milano and at National Insitute for the Physic of Matter (INFM) in 2000-2001, got a Ph.D. in Materials Engineering in 2004. He is Full professor at the Deprtment of Energy of Politecnico di Milano since 2019. He has experience in the field of nanostructures and nanostructured materials, nanostructured metal oxides and carbon-based nanomaterials deposition of thin films, scanning probe techniques (atomic force microscopy and scanning tunneling microscopy), inelastic light scattering (Raman, SERS and Brillouin even in situ). He published more than 100 papers. |